\documentclass[showpacs,reprint,superscriptaddress]{revtex4-1}

\usepackage{color}
\usepackage{graphicx}
\usepackage{hyperref}
\usepackage{amsmath}
\usepackage{array}
\begin{document}
\title{Dark Matter Search Results from the Commissioning Run of PandaX-II}
\date{\today}
\affiliation{INPAC and Department of Physics and Astronomy, Shanghai Jiao Tong University, Shanghai Laboratory for Particle Physics and Cosmology, Shanghai 200240, China}
\author{Andi Tan}
\affiliation{Department of Physics, University of Maryland, College Park, Maryland 20742, USA}

\author{Xiang Xiao}
\author{Xiangyi Cui}
\author{Xun Chen}
\affiliation{INPAC and Department of Physics and Astronomy, Shanghai Jiao Tong University, Shanghai Laboratory for Particle Physics and Cosmology, Shanghai 200240, China}
\author{Yunhua Chen}
\affiliation{Yalong River Hydropower Development Company, Ltd., 288 Shuanglin Road, Chengdu 610051, China}
\author{Deqing Fang}
\affiliation{Shanghai Institute of Applied Physics, Chinese Academy of Sciences, 201800, Shanghai, China}
\author{Changbo Fu}
\author{Karl Giboni}
\affiliation{INPAC and Department of Physics and Astronomy, Shanghai Jiao Tong University, Shanghai Laboratory for Particle Physics and Cosmology, Shanghai 200240, China}
\author{Franco Giuliani}
\affiliation{INPAC and Department of Physics and Astronomy, Shanghai Jiao Tong University, Shanghai Laboratory for Particle Physics and Cosmology, Shanghai 200240, China}
\affiliation{Center of High Energy Physics, Peking University, Beijing 100871, China}
\author{Haowei Gong}
\affiliation{INPAC and Department of Physics and Astronomy, Shanghai Jiao Tong University, Shanghai Laboratory for Particle Physics and Cosmology, Shanghai 200240, China}
\author{Shouyang Hu}
\affiliation{China Institute of Atomic Energy, Beijing 102413, China}
\author{Xingtao Huang}
\affiliation{School of Physics and Key Laboratory of Particle Physics and Particle Irradiation (MOE), Shandong University, Jinan 250100, China}
\author{Xiangdong Ji}
\email[Spokesperson: ]{xdji@sjtu.edu.cn}
\affiliation{INPAC and Department of Physics and Astronomy, Shanghai Jiao Tong University, Shanghai Laboratory for Particle Physics and Cosmology, Shanghai 200240, China}
\affiliation{Center of High Energy Physics, Peking University, Beijing 100871, China}
\affiliation{Department of Physics, University of Maryland, College Park, Maryland 20742, USA}
\author{Yonglin Ju}
\affiliation{School of Mechanical Engineering, Shanghai Jiao Tong University, Shanghai 200240, China}

\author{Siao Lei}
\author{Shaoli Li}
\affiliation{INPAC and Department of Physics and Astronomy, Shanghai Jiao Tong University, Shanghai Laboratory for Particle Physics and Cosmology, Shanghai 200240, China}

\author{Xiaomei Li}
\affiliation{China Institute of Atomic Energy, Beijing 102413, China}

\author{Xinglong Li}
\affiliation{China Institute of Atomic Energy, Beijing 102413, China}

\author{Hao Liang}
\affiliation{China Institute of Atomic Energy, Beijing 102413, China}

\author{Qing Lin}
\thanks{Now at Department of Physics, Columbia University}
\affiliation{INPAC and Department of Physics and Astronomy, Shanghai Jiao Tong University, Shanghai Laboratory for Particle Physics and Cosmology, Shanghai 200240, China}

%\author{Jianbei Liu}
%\affiliation{Department of Modern Physics, University of Science and Technology of China, Hefei 230026, China}

\author{Huaxuan Liu}
\affiliation{School of Mechanical Engineering, Shanghai Jiao Tong University, Shanghai 200240, China}

\author{Jianglai Liu}
\email[Corresponding author: ]{jianglai.liu@sjtu.edu.cn}
\affiliation{INPAC and Department of Physics and Astronomy, Shanghai Jiao Tong University, Shanghai Laboratory for Particle Physics and Cosmology, Shanghai 200240, China}

\author{Wolfgang Lorenzon}
\affiliation{Department of Physics, University of Michigan, Ann Arbor, MI, 48109, USA}

\author{Yugang Ma}
\affiliation{Shanghai Institute of Applied Physics, Chinese Academy of Sciences, 201800, Shanghai, China}
\author{Yajun Mao}
\affiliation{School of Physics, Peking University, Beijing 100871, China}

\author{Kaixuan Ni}
\thanks{Now at Department of Physics, University of California, San Diego}
\affiliation{INPAC and Department of Physics and Astronomy, Shanghai Jiao Tong University, Shanghai Laboratory for Particle Physics and Cosmology, Shanghai 200240, China}

\author{Kirill Pushkin}
\affiliation{INPAC and Department of Physics and Astronomy, Shanghai Jiao Tong University, Shanghai Laboratory for Particle Physics and Cosmology, Shanghai 200240, China}
\affiliation{Department of Physics, University of Michigan, Ann Arbor, MI, 48109, USA}

\author{Xiangxiang Ren}
\affiliation{INPAC and Department of Physics and Astronomy, Shanghai Jiao Tong University, Shanghai Laboratory for Particle Physics and Cosmology, Shanghai 200240, China}

\author{Michael Schubnell}
\affiliation{Department of Physics, University of Michigan, Ann Arbor, MI, 48109, USA}

\author{Manbin Shen}
\affiliation{Yalong River Hydropower Development Company, Ltd., 288 Shuanglin Road, Chengdu 610051, China}

\author{Fang Shi}
\affiliation{INPAC and Department of Physics and Astronomy, Shanghai Jiao Tong University, Shanghai Laboratory for Particle Physics and Cosmology, Shanghai 200240, China}

\author{Scott Stephenson}
\affiliation{Department of Physics, University of Michigan, Ann Arbor, MI, 48109, USA}

\author{Hongwei Wang}
\affiliation{Shanghai Institute of Applied Physics, Chinese Academy of Sciences, 201800, Shanghai, China}

\author{Jiming Wang}
\affiliation{Yalong River Hydropower Development Company, Ltd., 288 Shuanglin Road, Chengdu 610051, China}

\author{Meng Wang}
\affiliation{School of Physics and Key Laboratory of Particle Physics and Particle Irradiation (MOE), Shandong University, Jinan 250100, China}

\author{Qiuhong Wang}
\affiliation{Shanghai Institute of Applied Physics, Chinese Academy of Sciences, 201800, Shanghai, China}

\author{Siguang Wang}
\affiliation{School of Physics, Peking University, Beijing 100871, China}

%\author{Xiaolian Wang}
%\affiliation{Department of Modern Physics, University of Science and Technology of China, Hefei 230026, China}

\author{Xuming Wang}
\affiliation{INPAC and Department of Physics and Astronomy, Shanghai Jiao Tong University, Shanghai Laboratory for Particle Physics and Cosmology, Shanghai 200240, China}

\author{Zhou Wang}
\affiliation{School of Mechanical Engineering, Shanghai Jiao Tong University, Shanghai 200240, China}

\author{Shiyong Wu}
\affiliation{Yalong River Hydropower Development Company, Ltd., 288 Shuanglin Road, Chengdu 610051, China}

\author{Mengjiao Xiao}
\author{Pengwei Xie}
\email[Corresponding author: ]{willandy@sjtu.edu.cn}
\affiliation{INPAC and Department of Physics and Astronomy, Shanghai Jiao Tong University, Shanghai Laboratory for Particle Physics and Cosmology, Shanghai 200240, China}

\author{Binbin Yan}
\affiliation{School of Physics and Key Laboratory of Particle Physics and Particle Irradiation (MOE), Shandong University, Jinan 250100, China}

\author{Yong Yang}
\affiliation{INPAC and Department of Physics and Astronomy, Shanghai Jiao Tong University, Shanghai Laboratory for Particle Physics and Cosmology, Shanghai 200240, China}

\author{Jianfeng Yue}
\affiliation{Yalong River Hydropower Development Company, Ltd., 288 Shuanglin Road, Chengdu 610051, China}

\author{Xionghui Zeng}
\affiliation{Yalong River Hydropower Development Company, Ltd., 288 Shuanglin Road, Chengdu 610051, China}

\author{Hongguang Zhang}
\affiliation{INPAC and Department of Physics and Astronomy, Shanghai Jiao Tong University, Shanghai Laboratory for Particle Physics and Cosmology, Shanghai 200240, China}

\author{Hua Zhang}
\affiliation{School of Mechanical Engineering, Shanghai Jiao Tong University, Shanghai 200240, China}
\author{Huanqiao Zhang}
\affiliation{China Institute of Atomic Energy, Beijing 102413, China}

\author{Tao Zhang}
\author{Li Zhao}
\affiliation{INPAC and Department of Physics and Astronomy, Shanghai Jiao Tong University, Shanghai Laboratory for Particle Physics and Cosmology, Shanghai 200240, China}

%\author{Zhengguo Zhao}
%\affiliation{Department of Modern Physics, University of Science and Technology of China, Hefei 230026, China}
\author{Jing Zhou}
\affiliation{China Institute of Atomic Energy, Beijing 102413, China}
\author{Xiaopeng Zhou}
\affiliation{School of Physics, Peking University, Beijing 100871, China}

\collaboration{PandaX-II Collaboration}
\begin{abstract}
  We present the results of a search for WIMPs from the commissioning run of the PandaX-II experiment located at
  the China Jinping underground Laboratory. A WIMP search data set with an exposure of 306$\times$19.1 kg-day was taken, while 
  its dominant $^{85}$Kr background was used as the electron recoil calibration. No WIMP candidates are identified,  
  and a 90\% upper limit is set on the spin-independent elastic WIMP-nucleon cross section with a lowest 
  excluded cross section of 2.97$\times$10$^{-45}$~cm$^2$ 
  at a WIMP mass of 44.7~GeV/c$^2$. 
\end{abstract}
\maketitle

\section{Introduction}
\label{sec:intro}
%The existence of dark matter within the Galaxy and the entire universe
%is strongly supported by cosmological and astronomical
%observations~\cite{Read:2014qva, Harvey:2015hha, Ade:2015lrj}.  
The particle physics nature of dark matter is one of most fundamental
scientific questions. The leading candidates, weakly interacting massive 
particles (WIMPs),
% are the leading candidate
%of the dark matter particles.
%, which arise naturally from the physics
%beyond the Standard Model~\cite{Jungman:1995df, Bertone:2004pz}.  
can be directly detected by looking
for WIMP-nucleus scattering events in deep underground
laboratories. In recent years, experiments using the so-called
dual-phase xenon techniques have been continuously pushing the
exclusion limits of the elastic WIMP-nucleon scattering cross
section~\cite{Angle:2011th,Akimov:2011tj,Aprile:2013doa,Akerib:2013tjd,Akerib:2015rjg}, 
into the parameter space predicted by various theoretical
models~\cite{Cushman:2013zza}.

The PandaX experiment located at China Jinping underground Lab
(CJPL)~\cite{Yu-Cheng:2013iaa} is a dual-phase xenon direct dark matter detection
experiment~\cite{Cao:2014jsa}. The first phase of the experiment,
PandaX-I, with a 120-kg sensitive liquid xenon (LXe) target, performed
the WIMP search in 2014 with a 54$\times$80.1~kg-day exposure,
producing a strong limit on the WIMP-nucleon cross section for a WIMP
mass of less than 10~GeV/c$^2$~\cite{Xiao:2014xyn, Xiao:2015psa}, 
strongly disfavoring all positive claims from other experiments~\cite{Bernabei:2008yi,Aalseth:2010vx,Angloher:2011uu,Agnese:2013rvf}.  The
construction and installation of the second stage of the PandaX
experiment, PandaX-II, with a half-ton scale LXe target, commenced
after PandaX-I. In 2015, a series of engineering runs were carried out
to test the new detector system.  This is the largest running
dual-phase xenon detector to-date.  A brief physics commissioning run
was taken from Nov. 21 to Dec. 14, 2015, without dedicated 
electron recoil calibration and with a strong $^{85}$Kr background, 
based on which we report a WIMP search with a 306$\times$19.1~kg-day exposure.
\section{The PandaX-II Experiment}
\label{sec:pandax-ii}
\begin{figure}
\centering
\includegraphics[height=0.25\textheight]{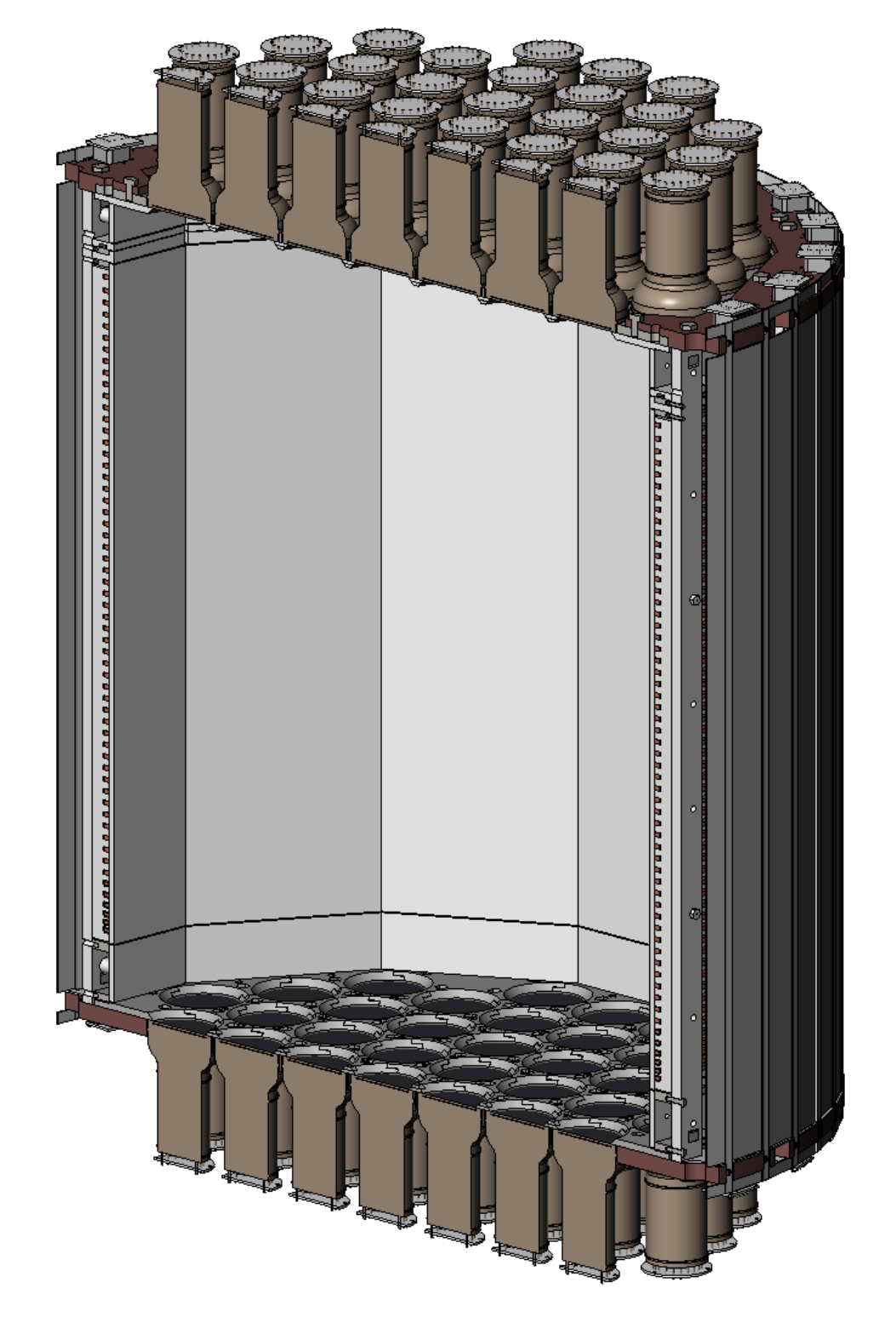}
\includegraphics[height=0.25\textheight]{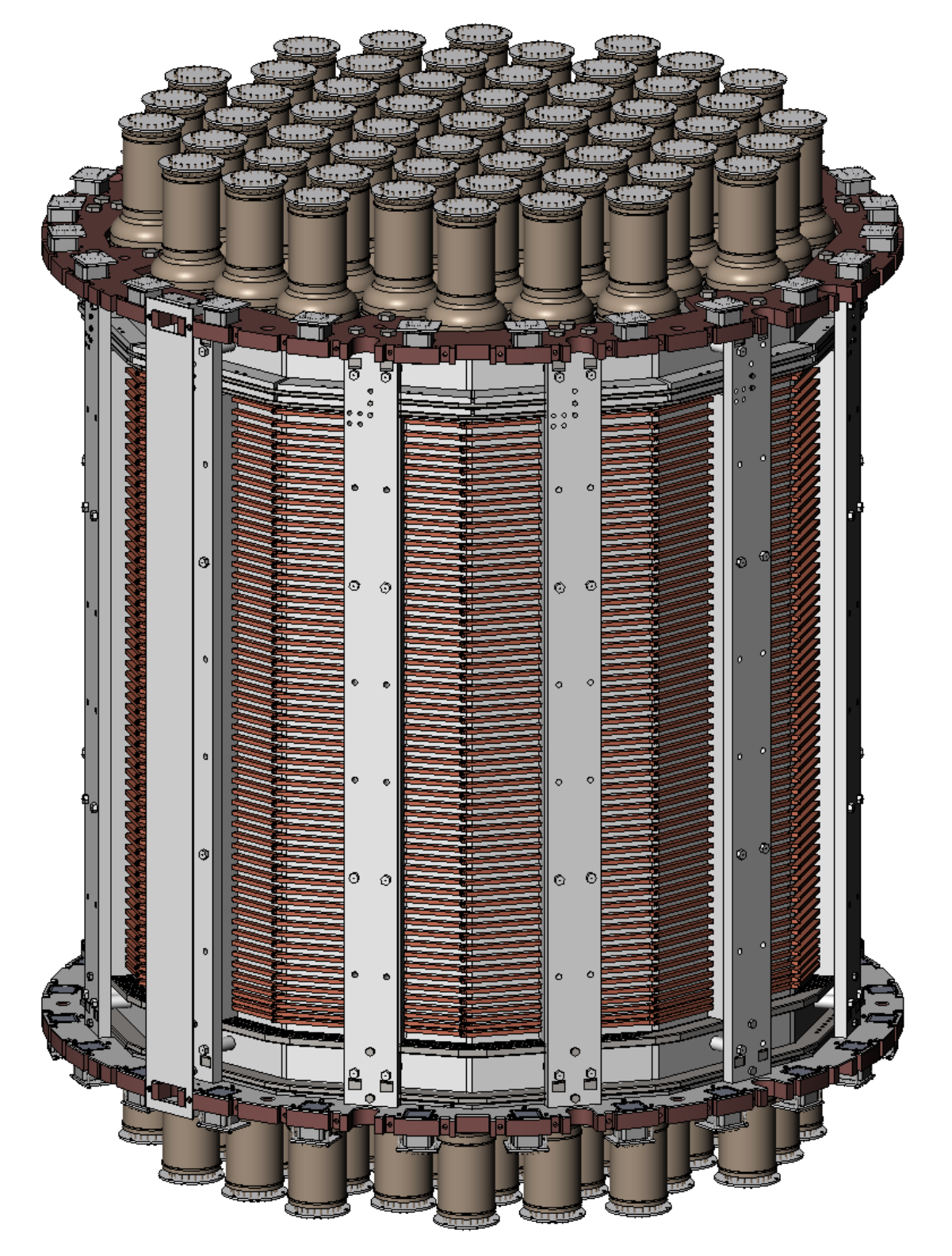}
\caption{Design drawings of PandaX-II TPC, left: cross-sectional view with both inner and outer PTFE panels, right: full view with skin volume exposed. See text for details.}
\label{fig:PandaXII_TPC}
\end{figure}

PandaX-II reuses most of the infrastructures of PandaX-I. The most
significant upgrades are the new inner vessel constructed from
stainless steel with much lower radioactivity, reducing the $^{60}$Co
activity by more than an order of magnitude, and a much larger xenon
Time Projection Chamber (TPC).  The cylindrical TPC, as shown in
Fig.~\ref{fig:PandaXII_TPC}, contains 580~kg LXe in the sensitive volume
enclosed by polytetrafluoroethylene (PTFE) reflective panels with an
inner diameter of 646~mm and a maximum drift length of 600~mm.  The
drift field is defined by a cathode mesh (200-$\mu$m wire diameter
with 5-mm pitch) placed at the bottom of the TPC and gate grid
(100-$\mu$m wire diameter with 5-mm pitch) 5.5~mm below the liquid
level. The liquid level can be adjusted remotely via an overflow mechanism.
The extraction field, which extracts electrons in liquid xenon into the gas region 
at the liquid-gas interface, is produced between the gate grid and the anode mesh
located 5.5~mm above the liquid level with the
same construction as the cathode.  
During the commissioning run, a voltage of $-$29~kV and $-$4.95~kV was
applied to the cathode and gate, respectively, and the anode was kept
at ground,
resulting in a drift field of 393.5 V/cm (with spatial variation of about 
0.77\% in the fiducial volume) in LXe, and an extraction ﬁeld of 4.4 kV/cm 
in the gaseous xenon right above the liquid surface.
Right outside the side
PTFE panels, 58 Cu shaping rings are mounted to guarantee the uniformity
of the drift field.  
A skin (surface layer) LXe volume with a thickness of about 40 mm 
is confined between the inner PTFE and a layer of outer PTFE panels.
Two
identical arrays of photomultiplier tubes (PMTs) were placed above and
below the TPC, respectively, each consisting of 55
Hamamatsu-R11410 3-inch PMTs, to detect scintillation photons in the
sensitive volume.  The top PMT array is placed 46~mm above the anode, 
and the bottom array is located 66~mm below the cathode.  A
screen grid (200-$\mu$m wire diameter with 5-mm pitch), set at ground,
is placed 6\,mm above the bottom PMT array to shield the cathode
high-voltage.  
Two additional PMT arrays are located at same heights as the 3-in arrays, 
each with 24 Hamamatsu-R8520-406 1-inch PMTs, to produce veto signals 
in the skin volume to suppress background events due to ambient gamma rays.
The PMT voltage divider for the 3-in PMTs 
uses a split positive and negative HV ($\sim\pm$650~V) scheme to reduce the relative potential
to the ground~\cite{Elsied:2015ixa}. The average random PMT rates (``dark count rate'')  
for the R11410 PMTs were measured 
to be $\sim$0.5 kHz, significantly improved from PandaX-I~\cite{Li:2015qhq}. 
%A Kapton-jacketed coaxial
%cable is used carry both the positive HV and signal, which gets
%decoupled by a 3-stage RC decoupler outside the detector. 
Same as that in PandaX-I, the signals from each PMT are amplified by a 
factor of 10, then get fed into the 100~MHz digitizer channels.

\section{Data Processing and Selection Cuts}
\label{sec:data_processing}
The same data acquisition setup from PandaX-I is used for
PandaX-II. Given the 60 cm depth of the TPC, the maximum time separation between the prompt 
photon signal in the liquid (S1) and the delayed proportional scintillation signal in the gas (S2) is estimated 
to be 350~$\mu$s. The length of each readout window 
is now 1~ms (200~$\mu$s in PandaX-I), with 500~$\mu$s before and after the trigger.   
The trigger is generated primarily on S2s for low energy events $<$10~keV$_{ee}$
electron equivalent energy~\cite{Ren:2016ium} with 
a trigger threshold of 79~photoelectrons (PE), and higher 
energy events are mostly triggered by S1s.

The data processing and signal selection followed the same framework
as in Ref.~\cite{Xiao:2015psa}, converting information from the raw waveforms 
of individual PMTs into photoelectrons and timing for S1s and S2s, vertex position, etc. 
The distortion of the mean waveform amplitude when there is no signal 
(baseline) induced by some large signals is corrected in software.
The single photoelectron gain (or PMT gain) is obtained by integrating the 
area of the waveform below the baseline for the single photoelectron signals. 
PMT gains are calibrated by activating LEDs inside the detector twice a week. 
The average PMT gain is $1.1\times10^{6}$.
After the gain correction, a threshold of 0.25 PE/sample in amplitude, roughly 
corresponding to a single channel threshold of 0.5 PE, is used for finding 
PMT hits from each waveform.
Clusters of time correlated hits are grouped into individual signals, which are 
tagged into either S1 signal, S2 signal or noise based on the shape of the 
summed waveform over all channels.
The discrimination between S1 and S2 signals relies on the 
full-width-10\%-maximum and the ``roughness'' of the waveform.  At least
three PMT hits are required for a valid S1 signal to suppress random coincidence 
among PMTs.
Veto PMTs hits are not used in the clustering. However any hit in the veto 
array that occurs during the entire width of an S1 signal, will veto an event.
The threshold to generate 
a veto was estimated to be $\sim$150~keV$_{ee}$ in the skin region
from a comparison between the data and Monte Carlo (MC) simulation.

On average the ratio of photoelectrons collected by the top and bottom
PMT arrays is $1:2$ for S1 and $2.2:1$ for S2. To suppress random S1-like
signals produced by the discharges on the electrodes or the so-called
``gamma-X'' events~\cite{Angle:2007uj} in the charge-insensitive regions, both likely to happen
close to the PMT arrays, selection cuts on the average
photoelectrons per-fired-PMT as well as the ratio of max-to-total
photoelectrons have been applied. Selection cuts are also set on the top-bottom ratio of S2
signals to remove spurious events located at the very edge and potential 
misidentified noises. In a given waveform,
the maximum number of S1-like signals passing all quality cuts is limited 
to two, and the maximum one is chosen to pair with S2.  To suppress events with 
incorrectly associated S1 and S2, the vertical
location encoded by the top-bottom asymmetry in S1 is required to be
consistent with that from the drift time. 
Finally, to avoid afterpulsing following an energetic event or discharge in the TPC, a
$>$10~ms separation between adjacent events is required in the dark
matter data.

The horizontal vertex position is reconstructed based on the charge pattern 
of S2 on the top PMT array. Like in PandaX-I, both a center-of-gravity and a
template matching reconstruction methods are used and cross checked. The average 
difference between the two is 10.8~mm within the fiducial volume (FV, defined later). This is a measure of the reconstruction
uncertainty, which leads to an 
uncertainty in the fiducial volume determination. 
The vertical position is obtained 
by the drift time, i.e. the time difference between S2 and S1, 
taking a drift speed of 1.7 mm/$\mu$s estimated from measured maximum drift time of 350~$\mu$s, 
also consistent with Ref.~\cite{Yoshino:1976zz} under a drift field of 
400 V/cm.
%The
%average difference between the two is required to be less than 40 mm.

\section{Detector Calibrations}
\label{sec:calibration}
To calibrate the detector response, a neutron source ($^{252}$Cf) and two $\gamma$ sources ($^{60}$Co and $^{137}$Cs)
were deployed through two PTFE tubes at different heights surrounding the inner vessel. Neutrons can excite xenon
nuclei or produce metastable nuclear states, leading to de-exciting $\gamma$ rays at 40 ($^{129}$Xe), 80 ($^{131}$Xe), 164 ($^{131m}$Xe), 
and 236 keV ($^{129m}$Xe). Photo-absorption $\gamma$ peaks
were used to calibrate the detector response. 
%The average 164 keV $\gamma$s rate was $\sim$0.06 mBq/kg during the 19.1 live-days. 
The 164~keV $\gamma$ events
were uniformly distributed in the detector and were used to produce a uniformity correction for the S1 and S2 signals. 
A 3-D correction map was produced for S1.
%, with the volume average being unity in the entire sensitive volume. 
For the S2 signals, the vertical uniformity correction was obtained by fitting S2 vs. the drift time using an exponential decay constant $\tau$, known 
as the electron lifetime. 
As expected, $\tau$ improved over time due to continuous xenon purification from 240$\mu$s to 552$\mu$s with an average of 324$\mu$s 
during the entire run. 
The S2 distribution in the horizontal plane was used to produce a 2-D correction map, independent of the drift time.

\begin{figure}
  \centering
  \includegraphics[width=0.45\textwidth]{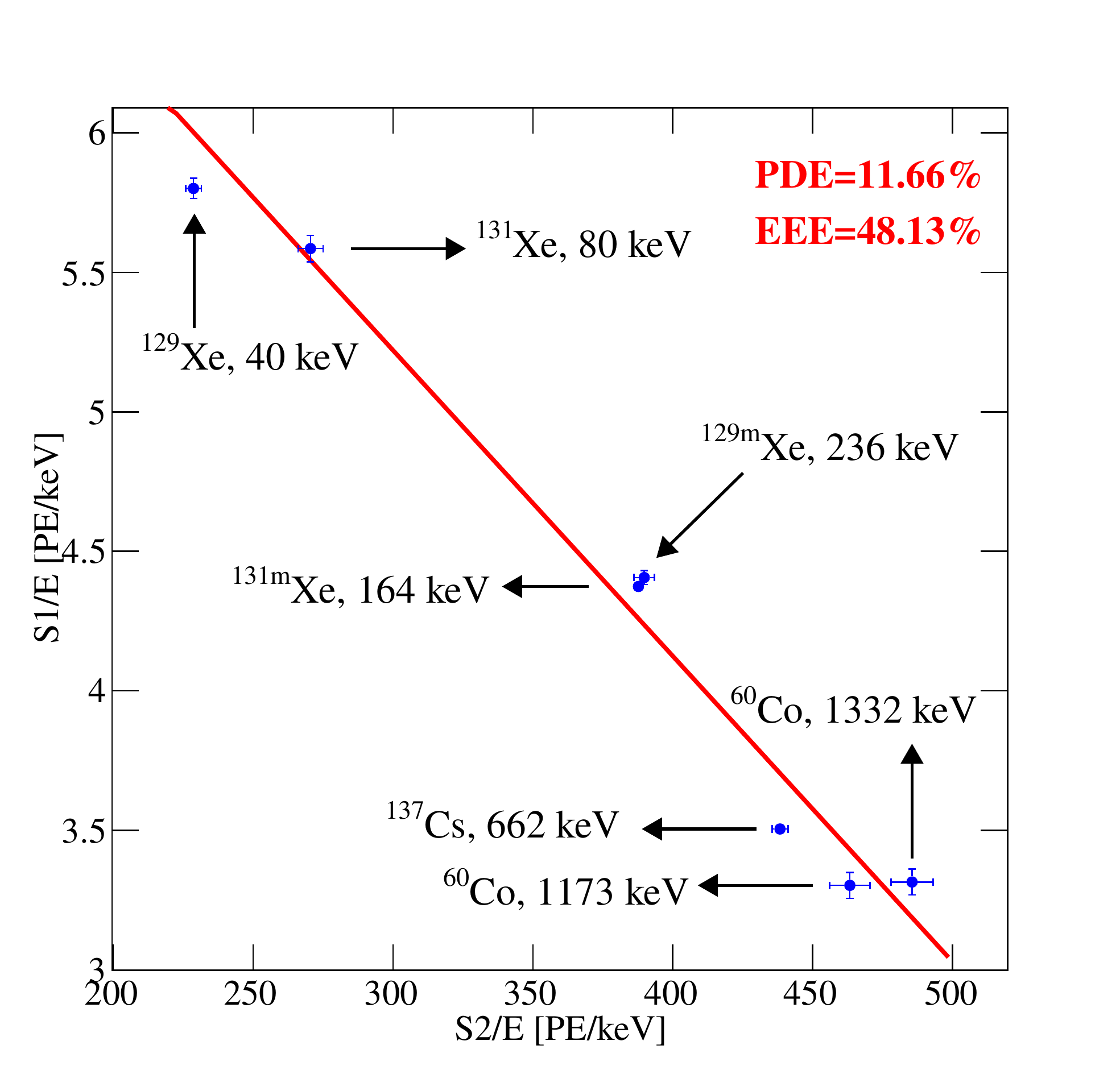}
  \caption{Linear fit in $\text{S2}/E_{ee}$ vs. $\text{S1}/E_{ee}$ for all $\gamma$ peaks. Each $\gamma$ peak was fit with a 2-D Gaussian in the 
    (S1, S2) plane with anti-correlation taken into account; only fit uncertainties are reflected on the data points.
  }
  \label{fig:doke_plot}
\end{figure}

The above uniformity correction was applied to all events. 
For each event, the electron equivalent energy $E_{ee}$ can be reconstructed as 
\begin{equation}
E_{ee} = W\times\left(\frac{\text{S1}}{\text{PDE}} + \frac{\text{S2}}{\text{EEE}\times\text{SEG}}\right)\,,
\end{equation}
in which $W = 13.7$~eV is the average work function to produce either an electron or photon~\cite{Lenardo:2014cva}. 
$\text{PDE}$, the photon-detection efficiency, $\text{EEE}$, the electron extraction efficiency, 
and $\text{SEG}$, the single-electron gain in PE/e, are the three key detector parameters to be 
determined. To obtain $\text{SEG}$, the smallest S2 signals in the data were identified as the single electron signals. Their 
photoelectron distribution was fit with a Gaussian function, from which SEG$=22.1\pm0.7$ with a resolution $\sigma=$7.41 PE/e was obtained.
To extract the other two parameters, the peak values of $\text{S1}/E_{ee}$ and $\text{S2}/E_{ee}$ 
are plotted for all $\gamma$ peaks (Fig.~\ref{fig:doke_plot}), 
where the true energy of the $\gamma$s are taken as $E_{ee}$. 
A linear fit is then performed on the data points. The scattering of data points along the line indicates 
systematic effects such as the non-uniformity and nonlinearity in S1 and S2. The best fit values,  
$\text{PDE} = 11.7\%, \text{EEE} = 48.1\%$, were compared with those obtained by taking the ratio of the observed peaks in S1 and S2 to the expected 
yield of photons and electrons from the NEST model~\cite{Lenardo:2014cva}. The difference leads to a relative 
uncertainty of 5.6\% in PDE and 7.1\% in EEE.

\begin{figure}[h]
  \centering
  \includegraphics[width=0.45\textwidth]{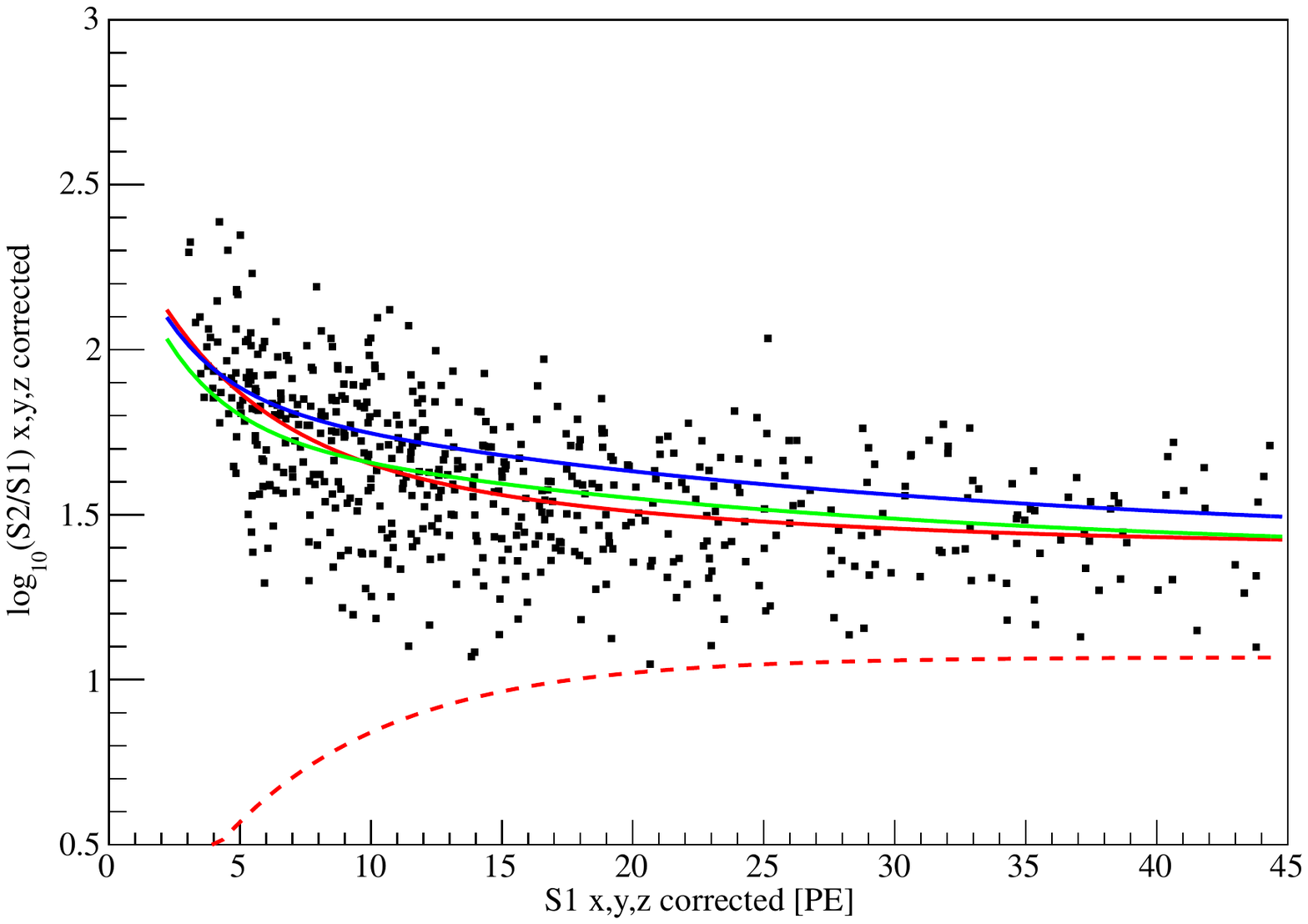}
  \caption{The comparison of $^{252}$Cf neutron calibration data with the median from the original NEST model (blue line) and the tuned 
    model (green line). A fit to the median of the data is given by the red line. 
    The dashed red line is the NR 99.99\% acceptance line based on the tuned MC 
    model (without cuts and efficiency applied) to remove spurious ``neutron-X'' 
    events with suppressed S2s due to charge loss in the inactive region.
}
  \label{fig:nest_prediction}
\end{figure}

The detector response to the low energy nuclear recoil (NR) events was
calibrated using $^{252}$Cf data.  The $\log_{10}(\text{S2/S1})$
vs. S1 of the NR events is shown in Fig.~\ref{fig:nest_prediction}. In total, 547 NR
events are identified for S1 between 3 to 45 PE 
in the FV.  The
MC predictions of the NR signal distribution were obtained from a
combination of Geant4-based program and the NEST model
with the extracted PDE, EEE, and SEG from calibration.  In simulating
photoelectrons, results from Ref.~\cite{Faham:2015kqa} were used to
incorporate double photoelectron emissions from the 3-in PMTs.  Vertical uniformity 
in S2 due to electron lifetime in the data and the S2 trigger threshold were 
also considered in the MC. The
median value of the MC is compared to the data in
Fig.~\ref{fig:nest_prediction}. A much better agreement can be
achieved by tuning the ratio of the initial number of excitation and
ionization by a factor of 1.5 in NEST. The width of the NR
band in the tuned MC also agrees with the data.  Therefore, we adopted the
tuned MC as the default model to predict the WIMP NR distributions.
The NR efficiency was evaluated by a comparison between the data and MC 
on the event 2-D distribution in (S1, S2), leading to a parametrization
\begin{equation}
\label{eq:eff}
\epsilon = 0.94 \left[e^{-\frac{\text{S1}-6.21}{1.66}}+1\right]^{-1}\left[e^{-\frac{\text{S2}_{\text{raw}}-79.3}{20.8}}+1\right]^{-1}\,,
\end{equation}
where $\text{S2}_{\text{raw}}$ is the raw S2 before the electron lifetime correction.
The energy independent factor, 0.94, was obtained by choosing high energy 
NR events with S1$>$20 PE and within $\pm3\sigma$ of the NR band, and 
removing the selection cuts. We identified two major effects which accounted 
for the efficiency loss at low recoil energy. First, due to the high rate ($\sim$160 Hz) 
from the $^{252}$Cf calibration source, the efficiency is significantly reduced by 
the presence of random single electron S2 signals. If the real S1 is small 
and the single electron S2 is mis-identified into multiple small S1 signals, the 
S1-S2 pairing algorithm would be ineffective and the event would be vetoed. 
In addition, the loss due to the three-fold PMT coincidence was also found to 
be significant due to multi-photoelectron emission in the R11410 
phototubes~\cite{Faham:2015kqa}. 
The NR efficiency in Eqn.~\ref{eq:eff} is conservatively taken as the dark matter 
detection efficiency.

In contrast, the NEST electron recoil (ER) model~\footnote{from the spreadsheet at \url{http://lux.physics.ucdavis.edu/~szydagis/NEST_LXe_LYandQYields_VsFieldParticle_Calc_Site.xlsx}} appears to describe the low energy data 
in the $\gamma$ calibration and WIMP search runs. However, inside the FV, the statistics of 
the low energy events in $\gamma$ calibration runs were insufficient due to the self-shielding effect from LXe. We shall resort to 
the WIMP search data to model the ER distribution.

\section{Backgrounds}
\label{sec:backgrounds}
As in Ref.~\cite{Xiao:2015psa}, the background is categorized into three components: the ER, neutron, and accidental background.

The ER background consists of external background due to
radioactivities in the detector materials, and internal backgrounds
due to krypton and radon. Before the detector assembly, the components
were assayed with a high purity Germanium counting station at CJPL. The assay results,
tuned to fit the data in high energy, were used as the input of the
Geant4-based MC to calculate the external ER background at low energy.
In the FV, such background is expected to be 0.21 mDRU (1~mDRU =
$10^{-3}$~evts$/$keV$/$kg$/$day).

In our data, significant number of low energy ER events were found 
and they distributed uniformly in the detector. They were identified as $^{85}$Kr $\beta$-decay
events (half-life 10.72 y) with a 99.563\% probability of single $\beta$ emission and
0.434\% $\beta-\gamma$ delayed-coincident emission. The krypton 
in xenon was likely introduced by an air leak during the previous 
fill and recuperation cycle. By fitting the
low energy ER events in the dark matter search data, the $^{85}$Kr is
estimated to be 0.082~mBq/kg or 15.04 mDRU with 3\% uncertainty. Assuming a concentration
of 2$\times10^{-11}$ in natural Kr, this leads to a Kr mole fraction
of 437$\pm$13 ppt in LXe, consistent with the offline gas sample measurement
using the technique from Ref.~\cite{Dobi:2011vc}. The $\beta-\gamma$ analysis
gave an independent estimate of 507$\pm$46 ppt Kr concentration. The difference
in the mean values from the two methods, or 17\%, is taken as the systematics 
uncertainty of the krypton background.

The radon level in LXe can be evaluated by identifying $\beta-\alpha$
and $\alpha-\alpha$ coincidence events. 
%or single $\alpha$s along the
%decay chain.  
$^{222}$Rn was estimated by the $^{214}$Bi-$^{214}$Po
events to be 6.57~$\mu$Bq/kg in the FV.  $^{220}$Rn was estimated by
the $^{212}$Bi-$^{212}$Po and $^{220}$Rn-$^{216}$Po events to be 0.54 and
0.41~$\mu$Bq/kg in the FV, respectively.  
%These values were about a
%factor of two less compared to those obtained using the single
%$\alpha$ events in the upper part of the chain, and the difference
%could be attributed to mobility of charged ions in electrical field
%and the liquid flow.  

Using MC, the contributions of low energy background discussed above 
are summarized in Table~\ref{tab:er_bkg_budget}.
%from Rn is summarized in Table~\ref{tab:er_bkg_budget},
%taking into account the secular in-equilibrium due to the long-lived
%$^{210}$Pb.

\begin{table}[h]
  \begin{tabular}{cc}
    \hline\hline
    Item & Background (mDRU) \\\hline
    Total & 15.33 \\
    $^{85}$Kr & 15.04 \\
    $^{222}$Rn & 0.075  \\
    $^{220}$Rn & 0.021  \\
%    $^{129m}$Xe, $^{131m}$Xe, $^{133}$Xe & $\sim$0 \\ 
    \hline
    PMT arrays \& bases & 0.097\\
    PTFE wall & 0.021 \\
    Inner vessel & 0.045 \\
    Others IV components & 0.026 \\
    Cu outer vessel & 0.016 \\
%    $^{238}$U & 0.068 \\
%    $^{232}$Th & 0.052 \\
%    $^{60}$Co & 0.049  \\
%    $^{40}$K & 0.015 \\
%    $^{137}$Cs & 0.004 \\
%    $^{235}$U & 0.001 \\
    \hline\hline
  \end{tabular}
  \caption{Summary of ER background from different components, including Rn, Kr, activated Xe, and other detector components. 
    The uncertainty is dominated by the $^{85}$Kr (17\%) based on two different analysis methods. }
  \label{tab:er_bkg_budget}
\end{table}

The neutron background can also be produced by the radioactivities of the detector components.
% by either 
%($\alpha$,n) reactions, spontaneous fission, or delayed neutron emission from decay of radionuclides. 
In our case, the PTFE material was measured to contain 3~mBq/kg $^{238}$U, of which the ($\alpha$,n) neutrons 
dominate the neutron background rate. Using the SOURCES-4A code~\cite{Wilson:1999so} with material radioactivities as inputs, 
the raw neutron rate and energy spectrum were determined. The neutron background was then calculated using the 
Geant4-NEST-based MC to be 0.06 events in the final data sets after all selection cuts, with a conservatively
estimated uncertainty of 100\%.
%based on which the NEST-based toy MC produces the background 
%distribution in (S1, S2). About 0.064 events passing all cuts could be found in the FV and 19.14 live-days.

The accidental background is produced by random coincidence of S1-like and S2-like signals. 
To evaluate it, dark matter search data were used to search for isolated S1 and S2 signals.
Since single small S1s are below the trigger threshold, they were searched in the 350~$\mu$s pre-trigger window 
of each event triggered by high energy S1. The rate was determined to be 2.8$\pm$0.1 Hz
within the S1 range cut.
%which is significant improvement compared to the $\sim$20~Hz rate observed in Ref.~\cite{Xiao:2015psa}. 
%However, within the 19.1-day DM run, 
In the same data set, 28069 single S2 events were identified within the final S2 range 
cut and radius cut.
%, a factor of 10 compared to that in Ref.~\cite{Xiao:2015psa}. 
%We also found significant rate (2500/day) of low energy events with S1 and S2 located close to the gate. These events, when 
%missing S1, would produce single S2 signals. 
When pairing the single S1 and S2 randomly in time, and with all coincidence selection 
cuts applied (which suppresses random events significiantly), 0.70 qualified accidental events are expected to survive
with a 25\% uncertainty estimated from the variation of S1 rates during the run.

The final expected background budget including ER, accidental, and neutron background is summarized in
Table~\ref{tab:backgroundtable}.
\begin{table}[h]
\centering
\begin{tabular}{cccc|c|c}
\hline\hline
 & ER & Accidental & Neutron & \parbox[t]{1.4cm}{Total\\Expected} & \parbox[t]{1.4cm}{Total\\observed}\\
\hline
All & 611 & 5.9 & 0.13 & 617$\pm$104 & 728 \\
\hline
\parbox{1.8cm}{Below \\NR median} & 2.5 & 0.7 & 0.06 & 3.2$\pm$0.71 & 2\\
\hline
\hline
\end{tabular}
\caption{The expected background events in 19.1 live-day dark matter search data in the FV, before and after the NR median cut. The uncertainties 
of the total expected background in the table are obtained based on the 17\%, 25\%, and 100\% uncertainties in the ER, accidental, and neutron 
background, respectively. Both the uncertainties from the ER rate (17\%) and leakage fraction (50\%) have been taken into account in that of the ER background below the NR median. See text for details. Number of events from the data are shown in the last column.}
\label{tab:backgroundtable}
\end{table}

\section{Final Candidates and WIMP Cross Section Limit}
\label{sec:events}
Only events with single S2 were selected into the final candidate set. 
The FV cut was determined to be within $r^2<$60000~mm$^2$ and
$20 \mu\text{s}<$drift time$<346\mu\text{s}$. The horizontal 
space facing the outermost ring of the 3-in PMTs is removed 
to avoid leakage from poorly reconstructed events, and the 
vertical cut is asymmetric since the bottom array has been
shielded by 66~mm of LXe. The vertex distributions in the data and MC 
are consistently flat within the FV. The amount of LXe in the cut is estimated to be 306$\pm$20~kg 
where the uncertainty arises from 
the 10.8~mm uncertainty in position reconstruction. 

A cut-based analysis was used to select dark matter candidate only from the 
events below the NR median curve from the data and above the 99.99\% NR acceptance curve from the NEST MC
 (shown in Fig.~\ref{fig:nest_prediction}).
To estimate the ER background leaking under the NR median curve (in the lack of dedicated ER calibration), we
used the distribution of the dark matter search data in the ($\log_{10}$(S2/S1), S1) plane
above the 33.3\%-NR-acceptance curve (1/3 of the NR events are located above it) for the NR events, performing Gaussian fits
to the data. The Gaussian leakage fraction was estimated to be
0.4$\pm$0.2\% below the NR median curve, confirmed by repeating the same estimate but including also data 
below the 33.3\%-NR-acceptance curve.
 
Based on the expected background, the final S1 range cut was chosen to be between 3 to 45 PE to 
give the optimal median sensitivity, corresponding to an average energy window between 1.3 to 8.7 keV$_{ee}$. 
S2s are required to be between 100 PE (raw) and 10000 PE (uniformity corrected). 
%balancing the background
%level and acceptance.

\begin{figure}
  \centering
  \includegraphics[width=0.45\textwidth]{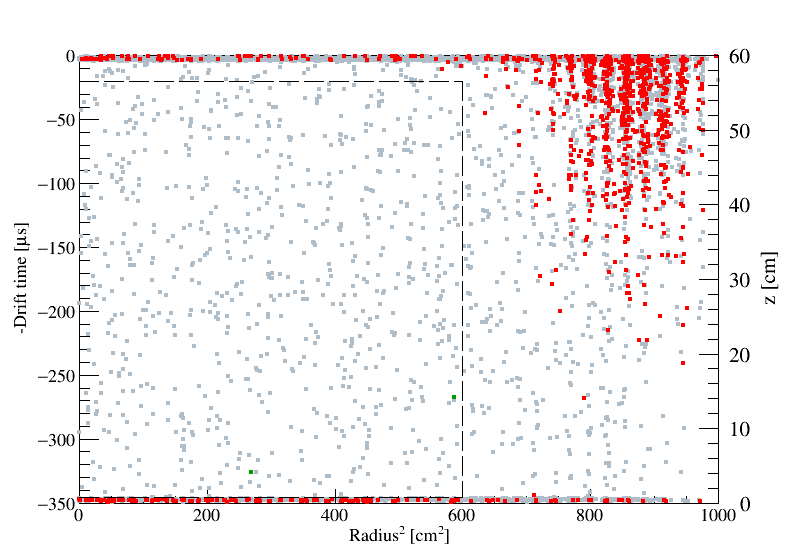}
  \caption{Position distribution of events that pass all selection cuts (gray points), and those below the NR median (outside FV: red points; inside FV: green stars), with FV cuts 
    indicated as the black dashed box. 
    The red points clustered at the top, bottom and upper right corner correspond 
    to events in these locations losing electrons on the electrodes or PTFE wall, 
    leading to a suppression of S2. The severe loss of S2 close to the bottom wall 
    leads to a significant event inefficiency indicated by the lack of events in the 
    lower right corner in the figure.
}
  \label{fig:events_pos}
\end{figure}

The event rates after various selection cuts are summarized in Table~\ref{tab:eventrate}. 
After the FV cut, 728 events are selected. 
The vertex distribution of all events before and after the FV cut
is shown in Fig.~\ref{fig:events_pos}. 
\begin{table}[h]
\centering
\begin{tabular}{ccc}
\hline\hline
Cut & \#Events & Rate (Hz) \\
\hline
All triggers & 4779083 & 2.89 \\
Single S2 cut & 1833756 & 1.11 \\
Quality cut & 1262906 & 0.76 \\
Skin veto cut & 1081044 & 0.65 \\
S1 range  & 45883 & 2.77 $\times 10^{-2}$ \\
S2 range  & 29755 & 1.80 $\times 10^{-2}$ \\
Fiducial volume & 728 & 4.40 $\times 10^{-4}$ \\
\hline\hline
\end{tabular}
\caption{The event rate in the dark matter runs after various analysis cuts.}
\label{tab:eventrate}
\end{table}

\begin{figure}[h]
  \centering
  \includegraphics[width=0.45\textwidth]{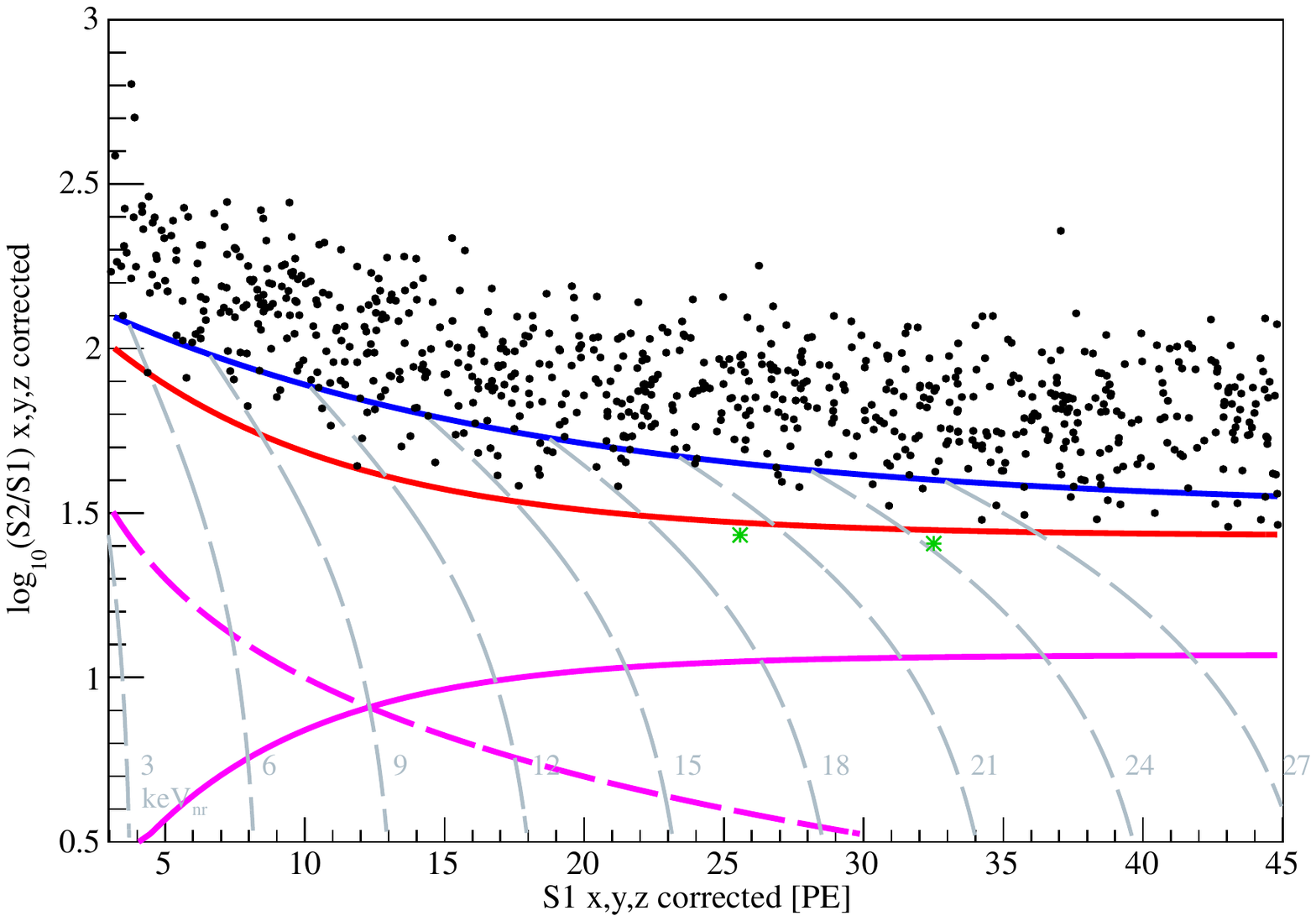}
  \caption{The distribution of $\log_{10}$(S2/S1) versus S1 for dark matter search data. The median of the NR calibration 
    band is indicated as the red curve. The dashed magenta curve is the equivalent 100~PE cut on S2. 
    The solid magenta and blue curves are the 99.99\% and 33.3\% NR acceptance curves, respectively. 
    The grey dashed curves are the equal energy curves 
    with NR energy indicated in the figures. The two data points located below the NR median curve are highlighted in green stars.}
  \label{fig:dm_band}
\end{figure}

The $\log_{10}$(S2/S1) vs. S1 distribution for the 728 candidates is shown in Fig.~\ref{fig:dm_band}. Two events were found just below the 
NR median curve, with their vertices indicated in Fig.~\ref{fig:events_pos}. 
Detailed examinations confirmed the high quality of these two events.
The distribution of $\log_{10}$(S2/S1) relative to the median of the data supports 
the assumption that band has a Gaussian profile in the vertical direction, 
which is applied to estimate the ER leakage fraction.

\begin{figure}[h]
  \centering
  \includegraphics[width=0.45\textwidth]{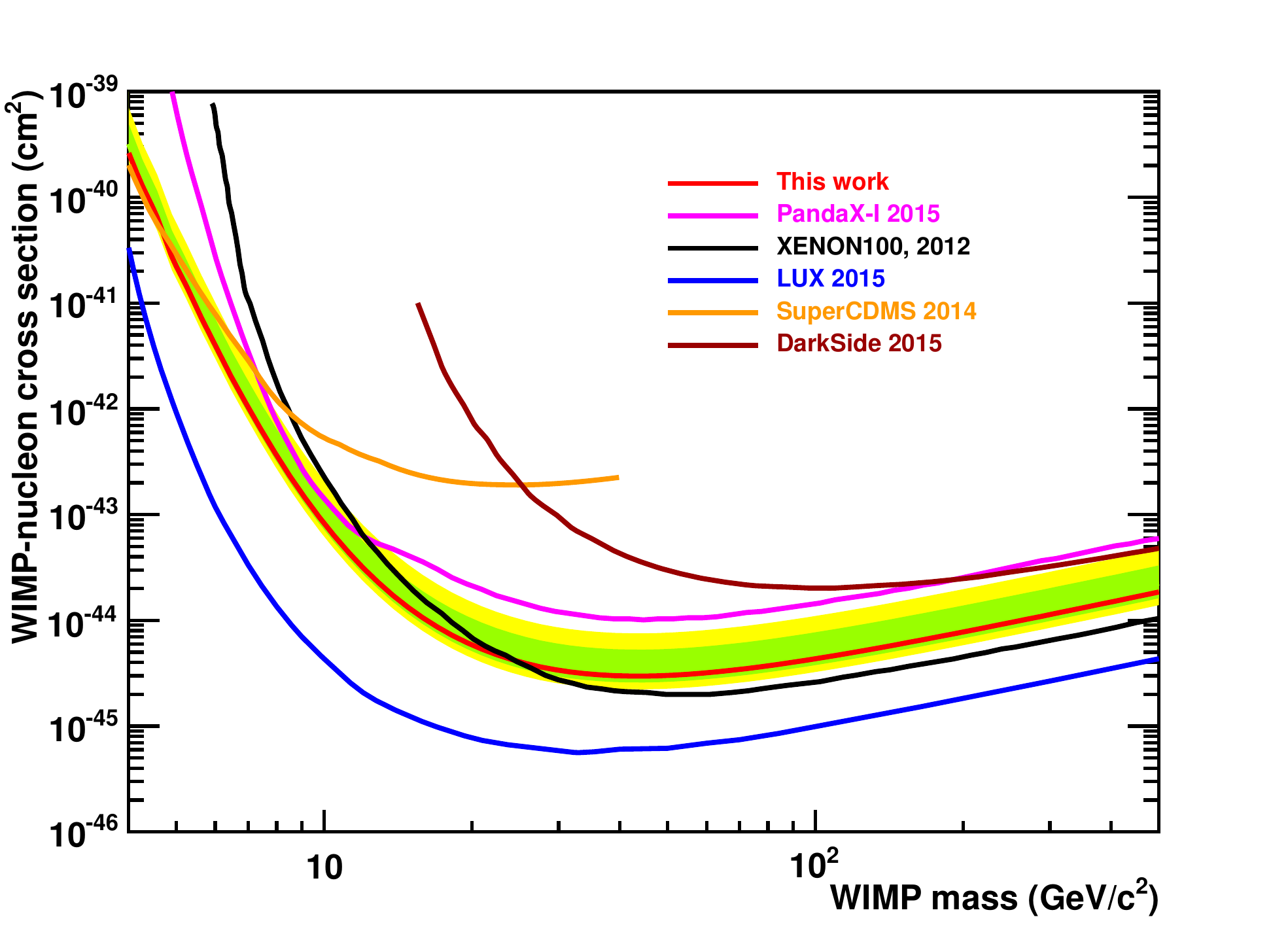}
  \caption{The 90\% C.L. upper limit for the spin-independent isoscalar WIMP-nucleon cross section from the PandaX-II commissioning run (red). 
    A selected set of recent world results are plotted for comparison: PandaX-I final results~\cite{Xiao:2015psa} (magenta), XENON100 225 day results~\cite{Aprile:2013doa} (black), LUX 2015 results~\cite{Akerib:2015rjg}(blue), SuperCDMS 2014~\cite{Agnese:2014aze}(orange), 
    and DarkSide 2015~\cite{Agnes:2015ftt} results (brown). The 
    $\pm1\sigma$ and $2\sigma$ sensitivity bands are shown in green and yellow, respectively.}
  \label{fig:limit}
\end{figure}

The final 90\% upper limit for the spin-independent isoscalar WIMP-nucleon cross section was calculated based on the two events 
and 3.2$\pm$0.71 expected background events (Table~\ref{tab:backgroundtable}) using the CLs method~\cite{CLS1,CLS2} with the same 
standard assumptions as in Ref.~\cite{Xiao:2015psa}. 
%The sensitivity bands were computed based on the expected background and its systematic uncertainty. 
The final results are shown in Fig.~\ref{fig:limit},  with recent results 
from PandaX-I~\cite{Xiao:2015psa}, XENON100~\cite{Aprile:2013doa}, LUX~\cite{Akerib:2015rjg}, SuperCDMS~\cite{Agnese:2014aze}, and DarkSide~\cite{Agnes:2015ftt} overlaid. 
Our upper limit lies within the $\pm$1$\sigma$ sensitivity 
band, and is more constraining due to the downward fluctuation of the background. 
In comparison, the upper limit computed based on the original NEST prediction
is weakened by about a factor of 2 in the high mass region but approaches a factor of 1.2 
at low mass ($\sim$10~GeV/c$^2$).
The lowest cross section limit obtained is 2.97$\times$10$^{-45}$~cm$^2$ at a WIMP mass of 
44.7 GeV/c$^2$, which represents an improvement of 
more than a factor of three from PandaX-I, and even more in the low WIMP mass region.
The major improvements include the exposure (a factor of $\sim$1.35), 
the photon detection (PDE 11.7\% vs. 9.6\%), the S2 selection cut (9.4 $e$ vs 19.7 $e$, although 
in PandaX-II a significant depth dependent efficiency variation is present due to the electron lifetime), 
the S1 window ([3, 45] vs. [2, 30] PE), and the expected background (3.2 vs. 6.9 events). 
The cross section limit at WIMP mass of 10, 100, and 300 GeV/c$^2$ are 
8.43$\times10^{-44}$, 4.34$\times10^{-45}$, and 1.13$\times10^{-44}$~cm$^2$, respectively. 
At low WIMP mass region down to 5~GeV/c$^2$, our exclusion limit is competitive with SuperCDMS~\cite{Agnese:2014aze}. 
At high WIMP mass region, our results are within a factor of $\sim$1.5 to the final 225-day XENON100 results~\cite{Aprile:2013doa}, although 
with only 19.1 days of live-time.
However, our results do not quite scale with the LUX 
results (with a factor of 2.4 of exposure)~\cite{Akerib:2015rjg} primarily due to the high 
krypton background.  

\section{Conclusions and Outlook}
\label{sec:conclusion}
In conclusion, we report the WIMP search results using the commissioning data of the 
PandaX-II experiment with an exposure of 306$\times$19.1 kg-day. No dark matter 
candidates were identified above background and 90\% upper limit is set on the 
spin-independent elastic WIMP-nucleon cross section with a lowest excluded 
value of 2.97$\times$10$^{-45}$~cm$^2$ at a WIMP mass of 44.7~GeV/c$^2$, 
a significant step-forward from PandaX-I.
After a brief maintenance period to distill krypton from xenon, the experiment is expected to 
resume physics data taking in spring 2016, and soon to explore previously unattainable 
WIMP parameter space. 

\begin{acknowledgments}
  This project has been supported by a 985-III grant from Shanghai
  Jiao Tong University, grants from National Science Foundation of
  China (Nos. 11435008, 11455001, 11505112 and 11525522), and a grant from the
  Office of Science and Technology in Shanghai Municipal Government
  (No. 11DZ2260700). This work is supported in part by the Chinese
  Academy of Sciences Center for Excellence in Particle Physics
  (CCEPP).  The project is also sponsored by Shandong University,
  Peking University, and the University of Maryland. 
  We also would like to thank Dr. Xunhua Yuan and Chunfa Yao of China Iron \& Steel Research Institute Group, 
  and we are particularly indebted to Director De Yin from Taiyuan Iron \& Steel (Group) Co. LTD 
  for crucial help on nuclear-grade steel plates. Finally, we thank
  the following organizations and personnel for indispensable
  logistics and other supports: the CJPL administration including
  directors Jianping Cheng and Kejun Kang and manager Jianmin Li, and
  the Yalong River Hydropower Development Company Ltd.
\end{acknowledgments}

\bibliography{refs.bib}
\end{document}